\documentclass[12pt,prd,aps,amssymb,amsmath,tightenlines,showkeys]{revtex4}

\usepackage{graphicx}

\newcommand{\half}{\textstyle{\frac{1}{2}}}

\newcommand{\cP}{{\cal P}}
\newcommand{\cT}{{\cal T}}

\begin{document}

\title{Periodic orbits for classical particles having complex energy}
\author{Alexander G. Anderson}\email{aganders@wustl.edu}
\author{Carl~M.~Bender}\email{cmb@wustl.edu}
\author{Uriel I.~Morone}\email{uimorone@wustl.edu}

\affiliation{Department of Physics, Washington University, St. Louis, MO 63130,
USA}

\date{\today}

\begin{abstract}
This paper revisits earlier work on complex classical mechanics in which it was
argued that when the energy of a classical particle in an analytic potential is
real, the particle trajectories are closed and periodic, but that when the
energy is complex, the classical trajectories are open. Here it is shown that
there is a discrete set of eigencurves in the complex-energy plane for which the
particle trajectories are closed and periodic.
\end{abstract}


\maketitle

\section{Introduction}
\label{s1}

This paper continues and advances the ongoing research program to extend
conventional real classical mechanics into the complex domain. Complex classical
mechanics is a rich and largely unexplored area of mathematical physics, and in
the current paper we report some new analytical and numerical discoveries.

In past papers on the nature of complex classical mechanics, many analytic
potentials were examined and some unexpected phenomena were discovered. For
example, in numerical studies it was shown that a complex-energy classical
particle in a double-well potential can exhibit tunneling-like behavior
\cite{R1}. Multiple-well potentials were also studied and it was found that a
complex-energy classical particle in a periodic potential can exhibit a kind of
band structure \cite{R2}. It was surprising to find classical-mechanical systems
that can exhibit behaviors that one would expect to be displayed only by
quantum-mechanical systems.

In previous numerical studies it was also found that the complex classical
trajectories of a particle having real energy are closed and periodic, but that
the classical trajectories of a particle having complex energy are generally
open. The claim that classical orbits are closed and periodic when the energy is
real and that classical orbits are open and nonperiodic when the energy is
complex was first made in Ref.~\cite{R3} and was examined numerically in
Ref.~\cite{R1}. In these papers it was emphasized that this property is
consistent with the Bohr-Sommerfeld quantization condition 
\begin{equation}
\oint_C dx\,p=\left(n+\half\right)\pi,
\label{e1}
\end{equation}
which can only be applied if the classical orbits are closed. Thus, there seems
to be an association between real energies and the existence of closed classical
trajectories \cite{R4}.

However, in this paper we show analytically that while a classical particle
having complex energy {\it almost always} follows an open and nonperiodic
trajectory, there is a special discrete set of curves in the complex-energy
plane for which the classical orbits are actually periodic. We call these curves
{\it eigencurves} because the requirement that the trajectory of a classical
particle having complex energy be closed and periodic is a kind of quantization
condition that specifies a countable set of curves in the complex-energy plane.
When the energy of a classical particle lies on an eigencurve, the trajectory of
the particle in the complex coordinate plane is periodic.

This paper is organized as follows: In Sec.~\ref{s2} we give a brief review of
complex classical mechanics focusing on the tunneling-like behavior of a
classical particle that has complex energy. In Sec.~\ref{s3} we describe the
special periodic orbits of a complex classical particle in a quartic double-well
potential. In Sec.~\ref{s4} we consider sextic and octic potentials. Finally, in
Sec.~\ref{s5} we make some brief concluding remarks.

\section{Background and Previous Results on Tunneling-like behavior in Complex
Classical Mechanics}
\label{s2}

For the past twelve years there has been an active research program to extend
quantum mechanics into the complex domain. Complex quantum mechanics has rapidly
developed into a rich and exciting area of physics. It has been found that if
the requirement that a Hamiltonian be Hermitian is weakened and broadened to
include complex non-Hermitian Hamiltonians that are $\cP\cT$ symmetric, some of
the quantum theories that result are physically acceptable because these
Hamiltonians possess two crucial features: (i) their eigenvalues are all real,
and (ii) they describe unitary time evolution. (A Hamiltonian is $\cP\cT$
symmetric if it is invariant under combined spatial reflection $\cP$ and time
reversal $\cT$ \cite{R5,R6}.) Such Hamiltonians have been observed in
laboratory experiments \cite{R7,R8}.

The study of complex classical mechanics arose in an effort to understand the
classical limit of complex quantum theories. In the study of complex classical
systems, the complex as well as the real solutions to Hamilton's differential
equations of motion are considered. In this generalization of conventional
classical mechanics, classical particles are not constrained to move along the
real axis and may travel through the complex plane.

Early work on the particle trajectories in complex classical mechanics is
reported in Refs.~\cite{R9,R10}. Subsequently, detailed studies of the complex
extensions of various one-dimensional conventional classical-mechanical systems
were undertaken: The remarkable properties of complex classical trajectories are
examined in Refs.~\cite{R11,R12,R13,R14,R15}. Higher dimensional complex
classical-mechanical systems, such as the Lotka-Volterra equations for
population dynamics and the Euler equations for rigid body rotation, are
discussed in Refs.~\cite{R3}. The complex $\cP\cT$-symmetric Korteweg-de Vries
equation has also been studied \cite{R16,R17,R18,R19,R20,R21,R22}.

In part, the motivation for extending classical mechanics into the complex
domain is that doing so might enhance one's understanding of the subtle
mathematical phenomena that real physical systems can exhibit. For example, some
of the complicated properties of chaotic systems become more transparent when
extended into the complex domain \cite{R23}. Second, studies of exceptional
points of complex systems have revealed interesting and potentially observable
effects \cite{R24,R25}. Third, recent work on the complex extension of quantum
probability density constitutes an advance in understanding the quantum
correspondence principle \cite{R26}. Fourth, and most relevant to the work in
this paper, is the prospect of understanding the nature of tunneling.

\begin{figure}
\begin{center}
\includegraphics[scale=0.3, viewport=0 0 1000 617]{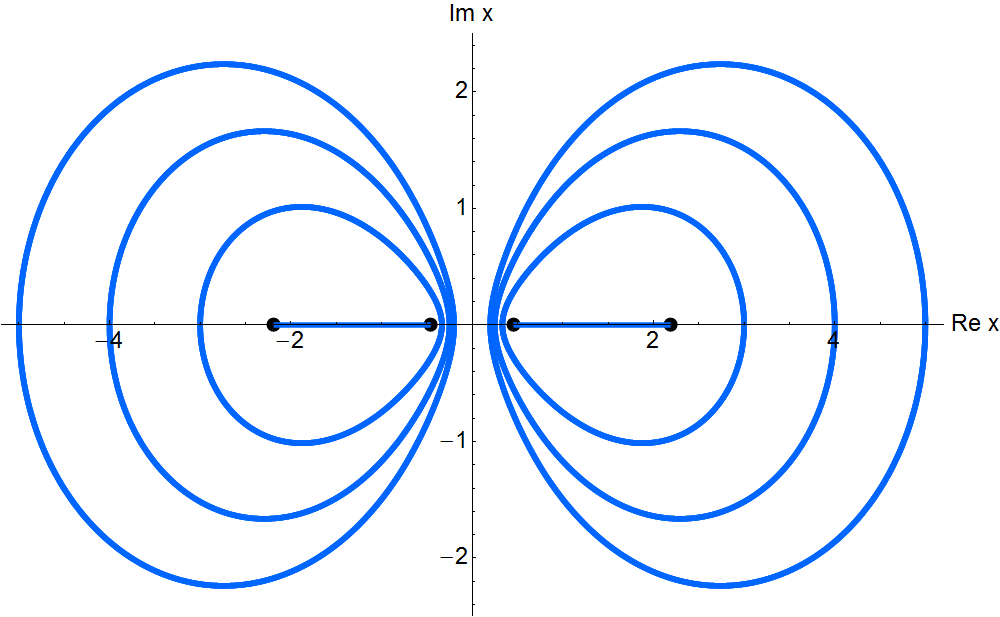}
\end{center}
\caption{Eight classical trajectories in the complex-$x$ plane representing a
particle of energy $E=-1$ in the potential $x^4-5x^2$. The turning points are
located at $x=\pm2.19$ and $x=\pm0.46$ and are indicated by dots. Because the
energy is real, the trajectories are all closed. The classical particle stays in
either the right-half or the left-half plane and cannot cross the imaginary
axis. Thus, when the energy is real, there is no effect analogous to tunneling.}
\label{F1}
\end{figure}

Let us illustrate how a classical particle can exhibit tunneling-like behavior.
Consider a classical particle in the quartic double-well potential $V(x)=x^4-5
x^2$. Figure \ref{F1} shows eight complex classical trajectories for a particle
of {\it real} energy $E=-1$. Each of these trajectories is closed and periodic.
Observe that for this energy the trajectories are localized either in the left
well or the right well and that no trajectory crosses from one side to the other
side of the imaginary axis.

\begin{figure}
\begin{center}
\includegraphics[scale=0.3, viewport=0 0 1000 976]{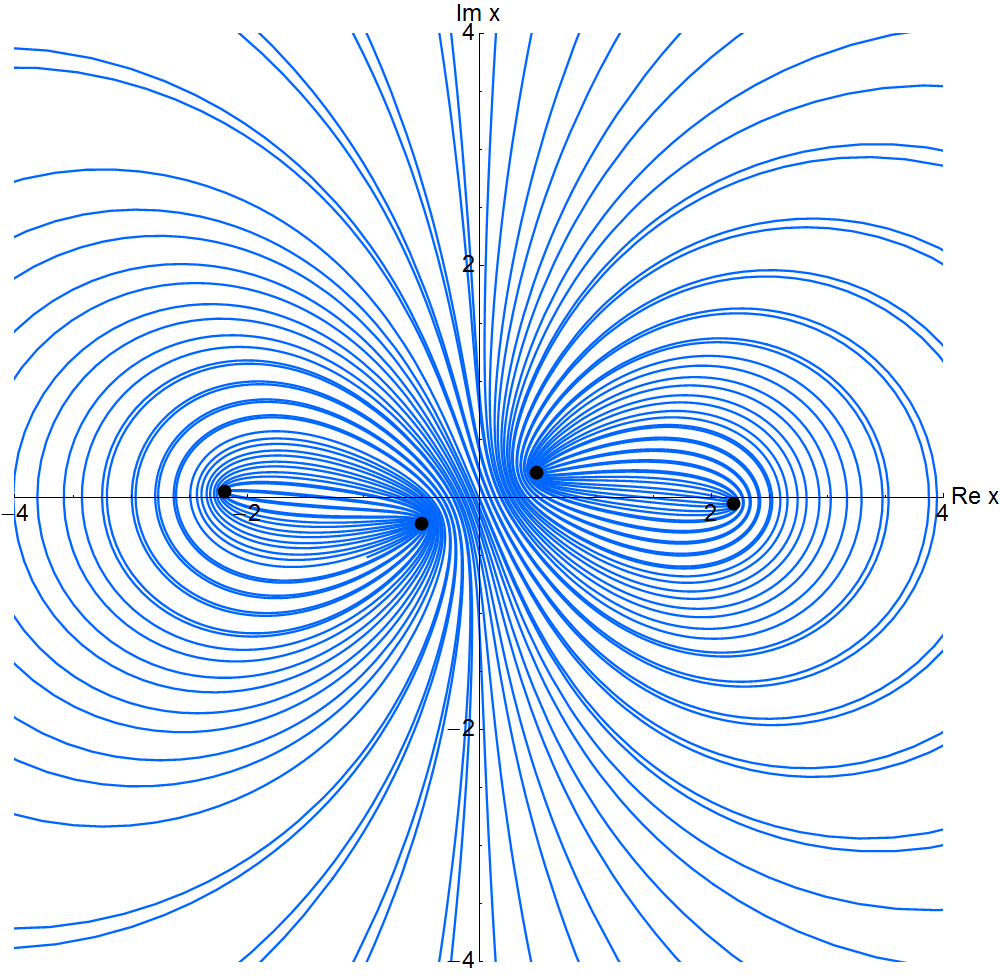}
\end{center}
\caption{Classical trajectory of a particle moving in the complex-$x$ plane
under the influence of a double-well $x^4-5x^2$ potential. The particle has
complex energy $E=-1-i$ and its trajectory does not close. The trajectory
spirals outward around one pair of turning points, crosses the imaginary axis,
and then spirals inward around the other pair of turning points. It then spirals
outward again, crosses the imaginary axis, and goes back to the original pair of
turning points. The particle repeats this behavior endlessly but at no point
does the trajectory cross itself. This classical-particle motion is analogous to
the behavior of a quantum particle that repeatedly tunnels between two
classically allowed regions. Here, the particle does not disappear into the
classically forbidden region during the tunneling process; rather, it moves
along a well-defined path in the complex-$x$ plane from one well to the other.}
\label{F2}
\end{figure}

What happens if we allow the classical energy to be complex \cite{R1}? In this
case the classical trajectory is generally not closed, but surprisingly it also
does not spiral out to infinity. Rather, the trajectory in Fig.~\ref{F2} unwinds
around a pair of turning points for a characteristic length of time and then
crosses the imaginary axis. At this point the trajectory does something
remarkable: Rather than continuing its outward journey, it spirals {\it inward}
towards the other pair of turning points. Then, never intersecting itself, the
trajectory turns outward again, and after the same characteristic length of
time, returns to the vicinity of the first pair of turning points. This
oscillatory behavior, which shares the qualitative characteristics of strange
attractors, continues forever but the trajectory never crosses itself. As in the
case of quantum tunneling, the particle spends a long time in proximity to a
given pair of turning points before rapidly crossing the imaginary axis to the
other pair of turning points. On average, the classical particle spends equal
amounts of time on either side of the imaginary axis. Interestingly, we find
that as the imaginary part of the classical energy increases, the characteristic
``tunneling'' time decreases in inverse proportion, just as one would expect of
a quantum particle.

The measurement of a quantum energy is inherently imprecise because of the
time-energy uncertainty principle $\Delta E\,\Delta t\gtrsim\hbar/2$.
Specifically, since there is not an infinite amount of time in which to make a
quantum energy measurement, the uncertainty in the energy $\Delta E$ is nonzero.
In Ref.~\cite{R1} the uncertainty principle was generalized to include the
possibility of {\it complex} uncertainty: If we suppose that the energy
uncertainty $\Delta E$ has a small imaginary component, then in the
corresponding classical theory, while the particle trajectories are almost
periodic, the orbits do not close exactly. The fact that the complex-energy
classical orbits are not closed means that in complex classical mechanics one
can observe tunneling-like phenomena that one normally expects to find only in
quantum systems.

\section{Periodic orbits in a quartic double-well potential}
\label{s3}

Let us consider the complex motion of a classical particle in the double-well
potential
\begin{equation}
V(x)=x^4-5x^2.
\label{e2}
\end{equation}
In general, the trajectory $x(t)$ of a classical particle in a potential $V(x)$
satisfies the differential equation
\begin{equation}
[x'(t)]^2+V(x)=E,
\label{e3}
\end{equation}
which is obtained by integrating Hamilton's equations of motion once. This
differential equation is separable, and for the double-well potential in
(\ref{e2}) the equation may be written as
\begin{equation}
dt=\frac{dx}{\sqrt{E+5x^2-x^4}}.
\label{e4}
\end{equation}

Integrating both sides of (\ref{e4}) gives rise to a Jacobi elliptic function.
The standard Jacobi elliptic function ${\rm sn}(u,k)$ is defined implicitly in
terms of the integral:
\begin{equation}
u=\int_0^{{\rm sn}(u,k)}\frac{ds}{\sqrt{(1-s^2)(1-k^2s^2)}}.
\label{e5}
\end{equation}
It is well known that the Jacobi elliptic function is doubly periodic and
that it satisfies the double periodicity condition 
\begin{equation}
{\rm sn}[u+2mK(k)+2niK(k'),k]=(-1)^m{\rm sn}(u,k),
\label{e6}
\end{equation}
where $m$ and $n$ are integers, $K(k)$ is the complete elliptic integral
\begin{equation}
K(k)\equiv\int_0^1\frac{ds}{\sqrt{(1-s^2)(1-k^2s^2)}},
\label{e7}
\end{equation}
and $k'\equiv\sqrt{1-k^2}$.

To identify the value of $k$ for the particular differential equation in
(\ref{e4}), we must factor the polynomial $E+5x^2-x^4$. Note that this
polynomial has four roots, $x=\pm a$ and $x=\pm b$, and thus we can write the
polynomial in factored form as
\begin{equation}
E+5x^2-x^4=-(a^2-x^2)(b^2-x^2).
\label{e8}
\end{equation}
By comparing powers of $x$ in (\ref{e8}) we determine that
\begin{equation}
a^2=\half\left(5-\sqrt{25+4E}\right),\qquad b^2=\half\left(5+\sqrt{25+4E}
\right).
\label{e9}
\end{equation} 
Using the factorization in (\ref{e8}) and replacing $x$ by $ax$ in (\ref{e4}),
we obtain the result
\begin{equation}
ibt=\int\frac{dx}{\sqrt{(1-x^2)(1-a^2x^2/b^2)}},
\label{e10}
\end{equation}
from which we identify
\begin{equation}
k^2=\frac{a^2}{b^2}=\frac{5-\sqrt{25+4E}}{5+\sqrt{25+4E}}.
\label{e11}
\end{equation}
Thus, the complex trajectory $x(t)$ of a classical particle that has complex
energy $E$ and moves according to the double-well potential $V(x)=x^4-5x^2$ is
given by
\begin{equation}
x(t)=a\,{\rm sn}(ibt,k),
\label{e12}
\end{equation}
where the time $t$ is real and the particle is initially at the origin.

The Jacobi elliptic function in (\ref{e12}) is doubly periodic according to
(\ref{e6}), so the trajectory in (\ref{e12}) closes if we make the replacement
\begin{equation}
ibt\to ibt+4mK(k)+2niK(k').
\label{e13}
\end{equation}
Moreover, since $t$ is real, we may eliminate this parameter by dividing the
expression in (\ref{e13}) by $ib$ and taking the imaginary part. We conclude
that the condition for having a periodic orbit is 
\begin{equation}
\frac{n}{m}=\frac{{\rm Im}[2iK(k)/b]}{{\rm Im}[K(k')/b]}.
\label{e14}
\end{equation}
This is the classical quantization condition referred to earlier in
Sec.~\ref{s1}.

\begin{figure}
\begin{center}
\includegraphics[scale=0.6, viewport=0 0 500 333]{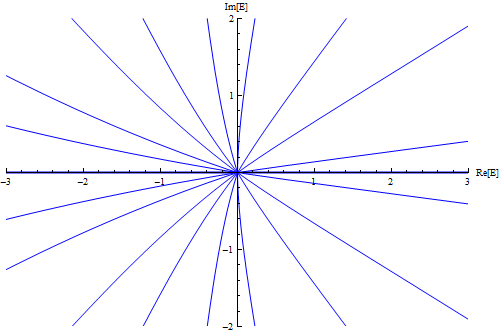}
\end{center}
\caption{Some quantized complex energies $E$ for the potential $V(x)=x^4-5x^2$
for $-3<{\rm Re}\,E<3$ and $-2<{\rm Im}\,E<2$ [see (\ref{e14})]. These curves
represent some of the (infinite number of) special complex energies $E$ for
which the classical orbits are periodic. These energies occur for rational
values of $n/m\geq2$. When $n=2$ and $m=1$, $E$ is real and positive. (This
corresponds to oscillatory particle motion above the barrier in the potential.)
The energy curve just above the positive-real axis in this figure corresponds to
$(n,m)=(5,2)$. Subsequent energy curves in anticlockwise order correspond to
$(n,m)=(3,1)$, $(n,m)=(4,1)$, $(n,m)=(5,1)$, $(n,m)=(7,1)$, $(n,m)=(10,1)$,
$(n,m)=(20,1)$, $(n,m)=(40,1)$, and the negative real axis corresponds to $n/m=
\infty$. (When $E<0$, the particle motion is oscillatory and confined to either
the left or the right well.) The energy curves in the lower-half $E$ plane are
complex conjugates of the energy curves in the upper-half $E$ plane. Near the
origin these curves are asymptotically straight lines [see (\ref{e15} and
(\ref{e16})].}
\label{F3}
\end{figure}

Equation (\ref{e14}) is an implicit equation for the complex number $E$ because,
as we can see from (\ref{e11}), the parameter $k$ (and hence $k'$) depends on
$E$. We are free to choose the integers $m$ and $n$, and once $m$ and $n$ are
chosen, (\ref{e14}) determines a countable set of eigencurves in the complex-$E$
plane; for any energy $E$ on these curves, the complex trajectory is periodic.
In fact, the complex trajectory for a complex energy satisfying (\ref{e14}) will
be periodic for {\it any} initial value $x(0)$. The curves in the complex-$E$
plane for which the particle trajectories $x(t)$ are periodic are shown in
Fig.~\ref{F3}. The curves in Fig.~\ref{F3} are symmetric with respect to the
real axis.

Let us examine (\ref{e14}) for small $|E|$ and ${\rm Im}\,E>0$. When $|E|<<1$,
$k^2\sim-\frac{2}{25}E$ and $k'\sim 1+E/25$. Thus, using the asymptotic
behaviors $K(\epsilon)\sim\half\pi$ ($\epsilon\to0$) and $K(1-\epsilon)\sim\log
(4/\sqrt{\epsilon})$ ($\epsilon\to0$), we see that when $0\leq{\rm arg}\,E\leq
\pi$, (\ref{e14}) reduces to 
\begin{equation}
{\rm arg}\,E\sim\pi\left(1-\frac{2m}{n}\right)\quad(|E|\to0).
\label{e15}
\end{equation}
In this formula $m$ and $n$ are integers with $n\geq2m$. When $-\pi\leq{\rm arg}
\,E\leq0$, the corresponding formula is
\begin{equation}
{\rm arg}\,E\sim-\pi\left(1-\frac{2m}{n}\right)\quad(|E|\to0).
\label{e16}
\end{equation}
Equations (\ref{e15}) and (\ref{e16}) show that the curves (see Fig.~\ref{F3})
for which the particle trajectories $x(t)$ are periodic emanate from $E=0$ as
straight lines before they begin to curve.

\begin{figure}
\begin{center}
\includegraphics[scale=0.6, viewport=0 0 500 357]{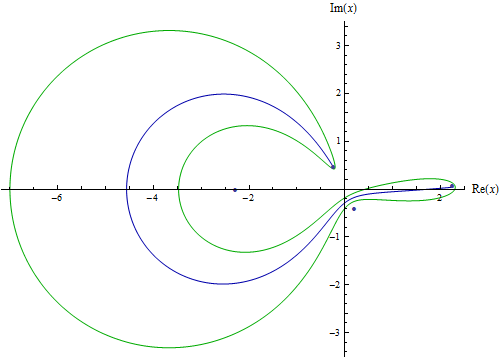}
\end{center}
\caption{Two periodic orbits for the quartic potential $V(x)=x^4-5x^2$
corresponding to $(n,m)=(3,1)$. We choose at random the complex energy $E=0.672
\,543\,108\,9+i$, which lies on the $(3,1)$ curve in Fig.~\ref{F3}. One orbit
(blue) oscillates between a pair of turning points (turning points are
indicated by dots) and the other orbit (green) encloses the pair of turning
points. Because of Cauchy's theorem all orbits having the same energy have
the same period.}
\label{F4}
\end{figure}

\begin{figure}
\begin{center}
\includegraphics[scale=0.6, viewport=0 0 500 373]{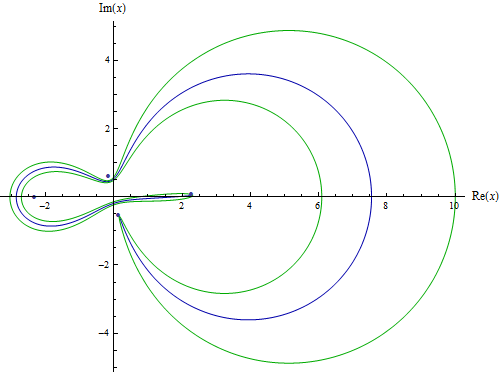}
\end{center}
\caption{Two periodic orbits for the potential $x^4-5x^2$ for the case $(n,m)=
(5,2)$. The energy is $E=1.540\,288\,094\,6+i$.}
\label{F5}
\end{figure}

\begin{figure}
\begin{center}
\includegraphics[scale=0.6, viewport=0 0 500 251]{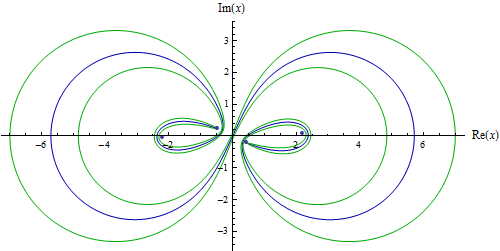}
\end{center}
\caption{Two periodic orbits for $V(x)=x^4-5x^2$ for the case $(n,m)=(8,1)$. The
energy is $E=-0.852\,958\,824\,6+i$.}
\label{F6}
\end{figure}

\begin{figure}
\begin{center}
\includegraphics[scale=0.6, viewport=0 0 500 268]{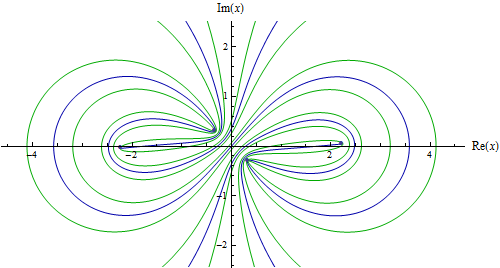}
\end{center}
\caption{Two periodic orbits for $V(x)=x^4-5x^2$ for the case $(n,m)=(14,3)$.
The energy is $E=-0.144\,984\,595\,5+i$.}
\label{F7}
\end{figure}

Some periodic orbits corresponding to the quantized energies are shown in
Figs.~\ref{F4} -- \ref{F7}. All curves for a given value of $n/m$ have the same
topology. In Fig.~\ref{F4} we display two periodic orbits corresponding to
$(n,m)=(3,1)$. For these curves we choose at random a complex energy $E=0.672\,
543\,108\,9+i$ that lies on the $(3,1)$ curve in Fig.~\ref{F3}. In Fig.~\ref{F5}
we take $(n,m)=(5,2)$ and choose the energy to be $E=1.540\,288\,094\,6+i$. In
Fig.~\ref{F6} we take $(n,m)=(8,1)$ and choose $E=-0.852\,958\,824\,6+i$. In
Fig.~\ref{F7} we take $(n,m)=(14,3)$ and choose $E=-0.144\,984\,595\,5+i$. 

Figures \ref{F4} -- \ref{F7} illustrate an easy way to determine the ratio $n/m$
by examining the shape of the orbits: First, one determines $m$ by counting the
number of times that the periodic orbit crosses the imaginary axis. Then one
determines $n$ by counting the number of times that the orbit crosses two
vertical lines, one between the left pair of turning points, and the other
between the right pair of turning points. For example, in Fig.~\ref{F4} the
blue orbit crosses the imaginary axis once, so $m=1$, and it crosses a
vertical line between the left pair of turning points twice and a vertical line
between the right pair of turning points once, so $n=3$. Thus, the ratio $n/m=
3$. For the green orbit we get $m=2$ and $n=6$, so again we find that $n/m=3$.

Similarly, in Fig.~\ref{F5} for the blue orbit we count $m=2$ and $n=5$ and for
the green orbit we count $m=4$ and $n=10$. Thus, $n/m=5/2$. In Figs.~\ref{F6}
and \ref{F7} it is easy to determine by inspection that $n/m=8$ and $n/m=14/3$.

\section{Sextic and Octic Potentials}
\label{s4}

\begin{figure}
\begin{center}
\includegraphics[scale=0.6, viewport=0 0 500 309]{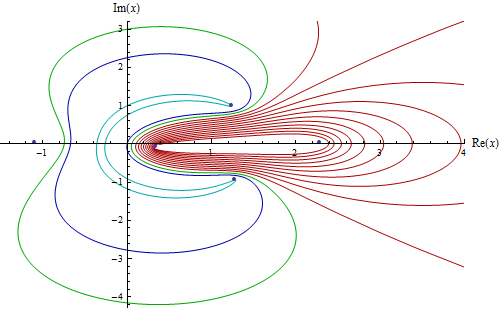}
\end{center}
\caption{Periodic and nonperiodic trajectories for the sextic potential in
(\ref{e17}). All trajectories correspond to the special complex energy $E=-4.359
\,375+i$ and are either periodic or nonperiodic depending on the initial
condition. The periodic orbits (blue, cyan, and green) are separated from the
nonperiodic trajectories (red) by a separatrix curve (not shown).}
\label{F8}
\end{figure}

Higher degree polynomial potentials are significantly more complicated than
quartic polynomial potentials because the classical trajectories are not
elliptic functions. Thus, it is very difficult to study such potentials
analytically. We have examined the complex trajectories of such potentials
numerically, and we have found that there are still special quantized complex
energy eigencurves for which the trajectories are periodic. However, the
remarkable feature of polynomial potentials having degree higher than four is
that now the behavior of trajectories depends on the initial condition. We find
that there is a separatrix in the complex-coordinate plane that divides the
periodic orbits from the nonperiodic paths. For example, in Fig.~\ref{F8}, which
displays some trajectories for the sextic potential
\begin{equation}
V(x)=x^6-5x^5-4x^4+11x^3-\textstyle{\frac{11}{4}}x^2-13x,
\label{e17}
\end{equation}
we have plotted four trajectories for the energy $E=-4.359\,375+i$. There are
three periodic orbits, one passing through $x=0$ (blue), a second passing
through $x=i$ (cyan), and a third passing through $x=3i$ (green). However, there
is a nonperiodic trajectory that begins at $x=5i$; this trajectory (red) spirals
inward in an anticlockwise direction around the pair of turning points that lie
just above and just below the positive real axis. Eventually this trajectory
will cross the midline joining these two turning points and will then spiral
outward. The nonperiodic trajectories are separated from the periodic
trajectories by a separatrix curve that crosses the imaginary axis near $4i$
(not shown).

\begin{figure}
\begin{center}
\includegraphics[scale=0.6, viewport=0 0 500 308]{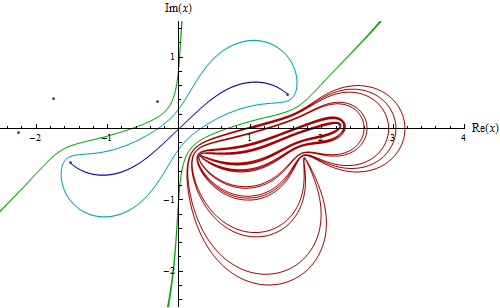}
\end{center}
\caption{Periodic and nonperiodic trajectories for the octic potential in
(\ref{e18}) for the energy $E=16.489+10i$. Two periodic trajectories (blue and
cyan) and a nonperiodic trajectory (red) is shown. Two separatrix curves (green)
separate the periodic trajectories from the nonperiodic trajectories. Numerical
calculations suggest that the two separatrix curves meet at infinity in the
first and third quadrants.}
\label{F9}
\end{figure}

In Fig.~\ref{F9} we plot some trajectories for the octic potential
\begin{equation}
V(x)=(x-1)^2(x+1)^2(x-2)^2(x+2)^2
\label{e18}
\end{equation}
for the special complex energy $E=16.489+10i$. There are two periodic orbits,
one starting at $x=0$ that oscillates between a pair of turning points (blue)
and a second (cyan) that passes through $x=0.3$ and encircles these turning
points. A nonperiodic trajectory (red) begins at $x=1$, spirals inward, then
outward, and then inward. As time passes, this trajectory will continue to
spiral outward and inward without ever crossing itself. The periodic and
nonperiodic trajectories are separated by a separatrix curve (green). One part
of the separatrix curve is shown passing through $x=0.670$ and the other part is
shown passing through $x=i$. These two curves join in the upper-right quadrant
and in the lower left quadrant. 

\section{Conclusions and Brief Remarks}
\label{s5}

We have shown in this paper that complex classical mechanics is richer and more
elaborate than previously imagined. While it is generally true that a classical
particle having complex energy traces an open trajectory, there are special
discrete quantized curves in the complex-energy plane for which the classical
particle has a periodic orbit.

For polynomial potentials the situation becomes more complicated as the degree
of the polynomial increases: For quartic potentials (which, according to the
lore of Riemann surfaces, are associated with the topology of a sphere) the
orbits are always periodic, regardless of whether the energy is real or complex.
For quadratic potentials (which are associated with the topology of a torus) the
trajectories are open except for a discrete set of eigencurves in the
complex-energy plane. When the energy lies on an eigencurve, the trajectory is
always periodic regardless of the initial condition. For sextic and octic
potentials (which are associated with the topology of a double and triple torus)
there are eigencurves for which, depending on the initial condition, the
particle trajectory may or may not be periodic. The periodic trajectories are
separated from the nonperiodic trajectories by a separatrix curve.

The behavior of complex classical trajectories is analogous to the behavior of a
quantum particle in a potential well. Ordinarily, a complex-energy classical
particle in a double well potential follows a space-filling spiral trajectory as
it alternately visits the left and the right well. However, we have shown in
this paper that there is a discrete set of complex energy eigencurves for which
the particle trajectories are periodic. The quantum analog is evident: The
initial wave function of a quantum particle in a double-well potential spreads
and diffuses as the particle tunnels from well to well. However, if the particle
is initially in an eigenstate, the wave function remains stationary and merely
oscillates in time.

One of our future objectives is to examine the nature of complex-energy
classical trajectories in periodic potentials. In the case of quartic potentials
one can make analytical progress because one can solve the equations of motion
in terms of elliptic functions. This is not in general possible for sextic and
higher-degree polynomial potentials. However, for periodic potentials such as
$V(x)=\sin x$, one can again solve the the equations of motion in terms of
elliptic functions. The behavior of complex trajectories in such potentials is
surprising, and we expect to complete a paper on this subject soon \cite{R2}.

\begin{acknowledgments} AGA and UIM are grateful to Washington University in
St.~Louis for partial support in the form of Undergraduate Research Fellowships.
This work was supported in part by a grant from the U.S.~Department of Energy.
Mathematica 7 was used to create the figures in this paper.
\end{acknowledgments}


\begin{thebibliography}{99}
\bibitem{R4} The harmonic oscillator is a glaring (and probably unique) example
of a classical dynamical system that has closed orbits even when the classical
energy is complex. When the classical energy is real, the classical orbits are
all ellipses whose axes are parallel to the real and the imaginary axes, and as
the energy becomes complex, the orientation of the ellipses changes.
\end{thebibliography}
\end{document}